\documentclass{elsarticle}
\usepackage{amsmath}
\usepackage{amssymb}

\usepackage{graphicx}
\usepackage{dcolumn}
\usepackage{bm}
\usepackage{cancel}
\usepackage{color}
\usepackage[utf8]{inputenc}

\begin{document}

\title{Diffusion on a lattice: transition rates, interactions and memory effects}

\author{M. A. Di Muro and M. Hoyuelos}
\ead{hoyuelos@mdp.edu.ar}
\address{Instituto de Investigaciones F\'isicas de Mar del Plata (IFIMAR-CONICET), Departamento de F\'isica, Facultad de Ciencias Exactas y Naturales, Universidad Nacional de Mar del Plata, Funes 3350, 7600 Mar del Plata, Argentina}

\begin{abstract}
We analyze diffusion of particles on a two dimensional square lattice. Each lattice site contains an arbitrary number of particles. Interactions affect particles only in the same site, and are macroscopically represented by the excess chemical potential. In a recent work, a general expression for transition rates between neighboring cells as functions of the excess chemical potential was derived. With transition rates, the mean field tracer diffusivity, $D^\text{MF}$, is immediately obtained. The tracer diffusivity, $D = D^\text{MF} f$, contains the correlation factor $f$, representing memory effects. An analysis of the joint probability of having given numbers of particles at different sites when a force is applied to a tagged particle allows an approximate expression for $f$ to be derived. The expression is applied to soft core interaction (different values for the maximum number of particles in a site are considered) and extended hard core.
\end{abstract}


\maketitle

\section{Introduction}

Diffusion and other transport processes have been widely studied due to
their applications to both academic research and industry; see, for example, \cite{peters,brodin,silva,cussler} and references cited therein.
A simple but yet useful approach to study diffusion are lattice gas models, in which atoms or particles occupy discrete positions in space. They are useful, as a first approximation, to characterize transport phenomena on regular structures, such as solids and surfaces \cite{ala,gomer,antczak,paul,mehrer}.
In these models particles jump to neighboring sites, albeit limited by the interaction with other particles. For instance, for hard-core interaction only one particle is permitted on each lattice site, and thus a particle is only allowed to jump to free-sites or vacancies.

Even with simple models, obtaining closed expressions for the
diffusivity for the complete range of density values has proved to be
challenging. One of the main issues are memory effects, which in
general manifest when the particle concentration becomes
significant. In the case of hard-core interaction, when a tagged
particle jumps to a free site, the empty site that it left behind is
bound to be the target of the next jump of the tagged particle, rather
than moving in other direction. Therefore, there is a spatial
correlation between consecutive jumps which causes the movement of a
tracer to deviate from a standard random walk.

In Ref.\ \cite{dimuro}, a formula for the jump rate in terms of the excess chemical potential was obtained, combining detailed balance with the Widom
insertion formula. The mean field (MF) regime is such that memory effects
can be neglected. Generally, the tracer diffusion coefficient is written in the form $D=D^\text{MF} f$, where $D^\text{MF}$ is the diffusion coefficient in the MF regime, and $f$ is known as the correlation factor, which takes
into account the memory effects discussed previously.

Much work (theoretical, numerical and experimental) has been devoted
to the determination of $f$ in different systems; references
\cite{compaan0,compaan,leclaire,howard,nakazato,kutner,kehr,tahir,chaturvedi,ferrando,hjelt,murch,mantina,bocquet,zhangdeng}
are just a small representative sample. Most applications are related
to diffusion in solids and, more specifically, to the vacancy
mechanism of self-diffusion or substitutional diffusion,
\cite[Ch.\ 7]{mehrer}.

Here we show that it is possible to develop an analytical expression for the correlation factor by taking into account averages of transition rates when a small force is applied to a tagged particle. We analyze a two-dimensional lattice, in which interactions are represented macroscopically by the excess chemical potential $\mu_{ex}$. Apart from hard-core, we focus on soft-core interaction, in which a site can contain up to $\Omega$ particles. Note that hard core is the special case of soft core with $\Omega=1$. In addition,  we analyze the case of extended hard core or $k$-NN (nearest-neighbor) hard core, in which the range of a particle occupying a site extends over a disk of a radius proportional to $k$. Extended hardcore is able to mimic, in the asymptotic limit of infinitely large $k$, the continuous dynamics of rigid disks via Monte Carlo simulations.

The paper is organized as follows. In Sec.\ \ref{s.corrfactor} we develop a general expression for the correlation factor using the transition rate formula for two neighboring sites. In Secs.\ \ref{s.softcore} and \ref{s.extended} we present expressions for the correlation factor for soft-core and extended hard-core interactions respectively. In addition, we compare the results predicted by the formulae with Monte Carlo simulations. Finally, in Sec.\ \ref{s.conclusions} we present the conclusions.

\section{Correlation factor}
\label{s.corrfactor}

Let us consider a $d$-dimensional square lattice. Each lattice site, identified with index $i$, is a cell of size $a$ containing an arbitrary number of particles. Interactions are considered only between particles within the same cell. There are $n_i$ particles in cell $i$. Each site is an open system at temperature $T$ and chemical potential $\mu$. We call $\Omega$ the number of microscopic states for one particle in a cell (proportional to its volume), and the (dimensionless) particle density is defined as $\rho_i = n_i/\Omega$.

It has been shown in Ref.\ \cite{dimuro} that the transition rate for one particle to jump from a cell with $n_1$ particles to a neighboring cell with $n_2$ particles is
\begin{equation}\label{e.trans}
W_{n_1,n_2} = \nu \frac{e^{-\beta (\mu_{\text{ex},n_2}-\mu_{\text{ex},n_1})/2}}{\sqrt{\Gamma_{n_1}\Gamma_{n_2}}},
\end{equation}
where $\nu$ is the jump frequency (a quantity that depends on the substratum and that is assumed constant), $\mu_{\text{ex},n_i}$ is the excess chemical potential and $\Gamma_{n_i}$ is the thermodynamic factor, defined as $\beta n_i \frac{\partial \mu}{\partial n_i} = 1+\beta n_i \frac{\partial \mu_{\text{ex},n_i}}{\partial n_i}$. The order of sub-indices in $W_{n_1,n_2}$ indicates the jump direction.

In a mean field approximation, we consider that $n_1=n_2=\bar{n}$ in Eq.\ \eqref{e.trans}, where $\bar{n}$ is the average number of particles. The resulting transition rate is $W_{\bar{n},\bar{n}}=\nu/\Gamma$, where $\Gamma$ without sub-index is evaluated at $\bar{n}$. The mean field tracer diffusivity is calculated using the continuous limit of a random walk with step $a$ \cite[Sect.\ 3.8.2]{gardiner}, giving 
\begin{equation}\label{e.DMF}
D^\text{MF} = W_{\bar{n},\bar{n}} a^2 = D_0/\Gamma,
\end{equation}
where $D_0= \nu a^2$ is the diffusivity at small concentration. It is known that this approximation for the tracer diffusivity is not appropriate in most cases \cite{ala}. Memory effects play an important role when considering, for example, hard core interaction; in this case, after a jump, a backward second jump is more likely than a forward jump since the origin site is empty. Memory effects are taken into account by the so-called correlation factor $f$, so that the tracer diffusivity is
\begin{equation}\label{e.Df}
D = D^\text{MF} f,
\end{equation}
or 
\begin{equation}\label{e.Df2}
\frac{D}{D_0} = \frac{1}{\Gamma} f.
\end{equation}
The correlation factor is usually written in terms of the average $\langle \cos\theta\rangle$, where $\theta$ is the angle between consecutive jump vectors (see for instance \cite[Ch.\ 7]{mehrer}). In the present approach we follow an alternative path appropriate for diffusion in a lattice that takes advantage from the expression for transition rates.

We consider diffusion of a tagged particle in a system composed by equivalent particles (in this case, tracer diffusivity is equivalent to self-diffusion coefficient). The tracer diffusion coefficient, $D$, can be obtained from the mobility, $B$, using the Einstein relation,
\begin{equation}\label{e.einstein}
D = \beta^{-1} B.
\end{equation}
When a small force, $F$, is applied to the tagged particle, its average velocity is $v = B F$; then, the tracer diffusivity is 
\begin{equation}\label{e.DvF}
D = \beta^{-1} v/F.
\end{equation}
Let us assume that the tagged particle is in site number 1, with $n_1$ particles. The velocity in terms of the transition rates is
\begin{equation}\label{e.vel}
v = a \langle W_{n_1,n_2}^F - W_{n_1,n_0}^F\rangle_F
\end{equation}
where 
\begin{align}
W_{n_1,n_2}^F &= W_{n_1,n_2} e^{\beta F a/2} = W_{n_1,n_2} (1 + \beta F a/2) \label{e.w12} \\
W_{n_1,n_0}^F &= W_{n_1,n_0} e^{-\beta F a/2} = W_{n_1,n_0} (1 - \beta F a/2) \label{e.w10}
\end{align}
and the average $\langle\rangle_F$ corresponds to a particle distribution with spatial correlations produced by the applied force. In equilibrium (without the external force), correlations are absent since interactions between neighboring sites are neglected. Replacing \eqref{e.w12} and \eqref{e.w10} in \eqref{e.vel}, we get
\begin{equation}\label{e.vel2}
v = a \langle W_{n_1,n_2} - W_{n_1,n_0}\rangle_F + \frac{\beta F a^2}{2} \langle W_{n_1,n_2} + W_{n_1,n_0}\rangle
\end{equation}
where the average $\langle\rangle$ in the last term corresponds to the equilibrium particle distribution with independent number probabilities, $P_{n_i}$, for each site. A small force is assumed in order to have a linear relationship between velocity and force.

Let us call $P^F_{n_0,n_1,n_2}$ the joint probability of having $n_0$, $n_1$ and $n_2$ particles in the corresponding sites when the force $F$ is applied to a tagged particle in site 1. The average of the first term in \eqref{e.vel2} is
\begin{equation}\label{e.averagef}
\langle W_{n_1,n_2} - W_{n_1,n_0}\rangle_F = \sum_{n_0,n_1,n_2} ( W_{n_1,n_2} - W_{n_1,n_0}) P^F_{n_0,n_1,n_2}.
\end{equation}
The joint probability is linearized,
\begin{equation}\label{e.lin}
P^F_{n_0,n_1,n_2} = P_{n_0}P_{n_1}P_{n_2} + \left.\frac{\partial P^F_{n_0,n_1,n_2}}{\partial F}\right|_{F=0} F,
\end{equation}
and the first derivative in $F$ is written in terms of an adimensional function $R_{n_0,n_1,n_2}$, that is defined such that
\begin{equation}\label{e.pr}
\left.\frac{\partial P^F_{n_0,n_1,n_2}}{\partial F}\right|_{F=0} = \beta a  P_{n_0}P_{n_1}P_{n_2} R_{n_0,n_1,n_2}.
\end{equation}
Then, 
\begin{equation}\label{e.lin2}
P^F_{n_0,n_1,n_2} = (1 + \beta F a R_{n_0,n_1,n_2}) P_{n_0}P_{n_1}P_{n_2}
\end{equation}
and the average in \eqref{e.averagef} is
\begin{equation}\label{e.aver2}
\langle W_{n_1,n_2} - W_{n_1,n_0}\rangle_F = \beta F a \langle (W_{n_1,n_2} - W_{n_1,n_0}) R_{n_0,n_1,n_2}\rangle,
\end{equation}
that is, we have rewritten a non-equilibrium average in terms of an equilibrium average using the function $R_{n_0,n_1,n_2}$ (also, it was used that, at equilibrium, $\langle W_{n_1,n_2}\rangle = \langle W_{n_1,n_0}\rangle$). Using \eqref{e.aver2} in \eqref{e.vel2}, the average velocity is
\begin{equation}\label{e.vel3}
v = \beta F a^2 \langle (W_{n_1,n_2} - W_{n_1,n_0}) R_{n_0,n_1,n_2} + W_{n_1,n_2}\rangle
\end{equation}
and, using \eqref{e.DvF}, the tracer diffusivity is
\begin{equation}\label{e.D}
\frac{D}{D_0} = \frac{1}{\nu} \langle (W_{n_1,n_2} - W_{n_1,n_0}) R_{n_0,n_1,n_2} + W_{n_1,n_2}\rangle.
\end{equation}
The correlation factor, $f$, is obtained combining this equation with \eqref{e.Df2}:
\begin{equation}\label{e.f}
f = \frac{\langle (W_{n_1,n_2} - W_{n_1,n_0}) R_{n_0,n_1,n_2} + W_{n_1,n_2}\rangle}{W_{\bar{n},\bar{n}}},
\end{equation}
where it was used that $W_{\bar{n},\bar{n}}=\nu/\Gamma$. It is shown below, with some examples, that simple approximations for $R_{n_0,n_1,n_2}$ provide appropriate expressions for the correlation factor.

\section{Soft core interaction}
\label{s.softcore}

For soft core interaction, the thermodynamic factor is $\Gamma=1/(1-\rho)$ and, replacing in \eqref{e.DMF}, the mean field tracer diffusivity is \cite{dimuro}
\begin{equation}\label{e.Dmfsoft}
D^\text{MF}/D_0 = 1 - \rho,
\end{equation}
where $\rho=\bar{n}/\Omega$, with $\bar{n}$ the average particle number in one cell and $\Omega$ the maximum number of particles allowed (a measure of the cell's volume). Hard core interaction is obtained for $\Omega=1$. The transition rate is \cite{dimuro}
\begin{equation}\label{e.W}
W_{n_1,n_2} = \nu (1-\rho_2)
\end{equation}
with $\rho_2=n_2/\Omega$. In the present notation, $\rho$ (without sub-index) is the average concentration at any point of the lattice, while $\rho_i=n_i/\Omega$ is the instantaneous concentration at site $i$.  The transition rate depends only on the particle number in the destination site. Using this information in Eq.\ \eqref{e.f}, the correlation factor is
\begin{equation}\label{e.fsc}
f = 1 - \frac{1}{1-\rho} \langle (\rho_2-\rho_0) R_{n_0,n_2} \rangle,
\end{equation}
where the dependence of $R$ on $n_1$ can be omitted since the averaged quantity does not depend on $n_1$; more explicitly, $R_{n_0,n_2}=\sum_{n_1} R_{n_0,n_1,n_2} P_{n_1}$.

An expansion in powers of the concentrations in sites 0 and 2 is proposed for $R_{n_0,n_2}$:
\begin{equation}\label{e.expR}
R_{n_0,n_2} = \sum_{i,j=0}^\infty c_{i,j} \rho_0^i \rho_2^j,
\end{equation}
where it is assumed that the coefficients $c_{i,j}$ do not depend on $\Omega$. Keeping terms up to order 2, the average involving $R_{n_0,n_2}$ in \eqref{e.fsc} is 
\begin{align}\label{e.averRsc}
\langle (\rho_2-\rho_0)& R_{n_0,n_2} \rangle \nonumber \\ 
 =& \langle (\rho_2-\rho_0) (c_{00} + c_{10}\rho_0 + c_{01}\rho_2 + c_{20}\rho_0^2 + c_{02}\rho_2^2 + c_{11} \rho_0\rho_2)\rangle \nonumber\\
 =& \frac{(c_{01}-c_{10})}{\Omega^2}(\langle n^2\rangle - \bar{n}^2) + \frac{(c_{02}-c_{20})}{\Omega^3} (\langle n^3\rangle - \bar{n} \langle n^2\rangle)\nonumber\\
 =& \frac{c'}{\Omega^2} \langle\Delta n^2\rangle + \frac{c}{\Omega^3}(\langle\Delta n^3\rangle + 2 \bar{n}\langle\Delta n^2\rangle),
\end{align}
where $c'= c_{01}-c_{10}$, $c=c_{02}-c_{20}$, and $n$ is used indistinctly for sites 0 or 2 since equilibrium averages are the same in both sites. The average particle number for soft-core interaction is given by the Fermi-Dirac distribution:
\begin{equation}\label{e.nmediosoft}
\bar{n} = \frac{\Omega}{1 + e^{-\beta \mu}}
\end{equation}
from which the second and third order moments are obtained: $\langle \Delta n^2\rangle = \frac{1}{\beta} \frac{\partial \bar{n}}{\partial \mu} = \bar{n}(1-\rho)$ and $\langle \Delta n^3\rangle=\frac{1}{\beta^2} \frac{\partial^2 \bar{n}}{\partial \mu^2} = \bar{n}(1-\rho)(1-2\rho)$. Using this information and going back to the expression for the correlation factor, Eq.\ \eqref{e.fsc}, we obtain,
\begin{equation}\label{e.fsc2}
f = 1 + \frac{\rho}{\Omega}[c'+ c/\Omega + c\, 2(1-1/\Omega)\rho].
\end{equation}
Compaan and Haven \cite{compaan} have demonstrated that, for hard core interaction ($\Omega=1$) in a two dimensional lattice and in the limit of $\rho\rightarrow 1$, the correlation factor takes the value $f_{\Omega=1} = 1/(\pi-1)\simeq 0.46694$. This result can be used to set one of the constants in \eqref{e.fsc2}, where $f_{\Omega=1} = 1 + c'+ c$. Then,
\begin{equation}\label{e.fsoft}
f = 1+ \frac{\rho}{\Omega} \left[ \frac{\pi-2}{\pi-1} + c \left(1-\frac{1}{\Omega}\right)(2\rho-1) \right]
\end{equation}

\begin{figure}
	\includegraphics[width=\linewidth]{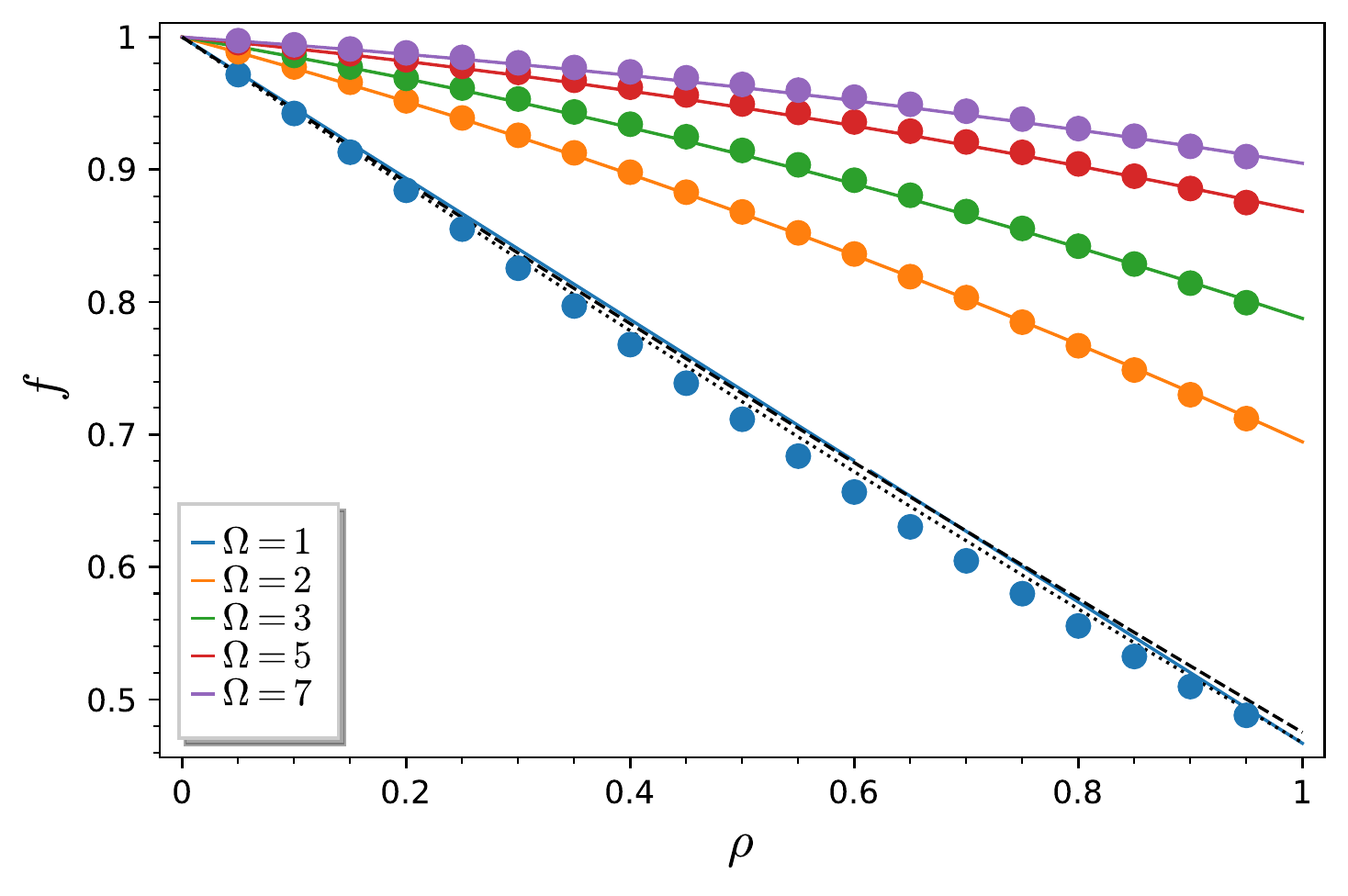}
	\includegraphics[width=\linewidth]{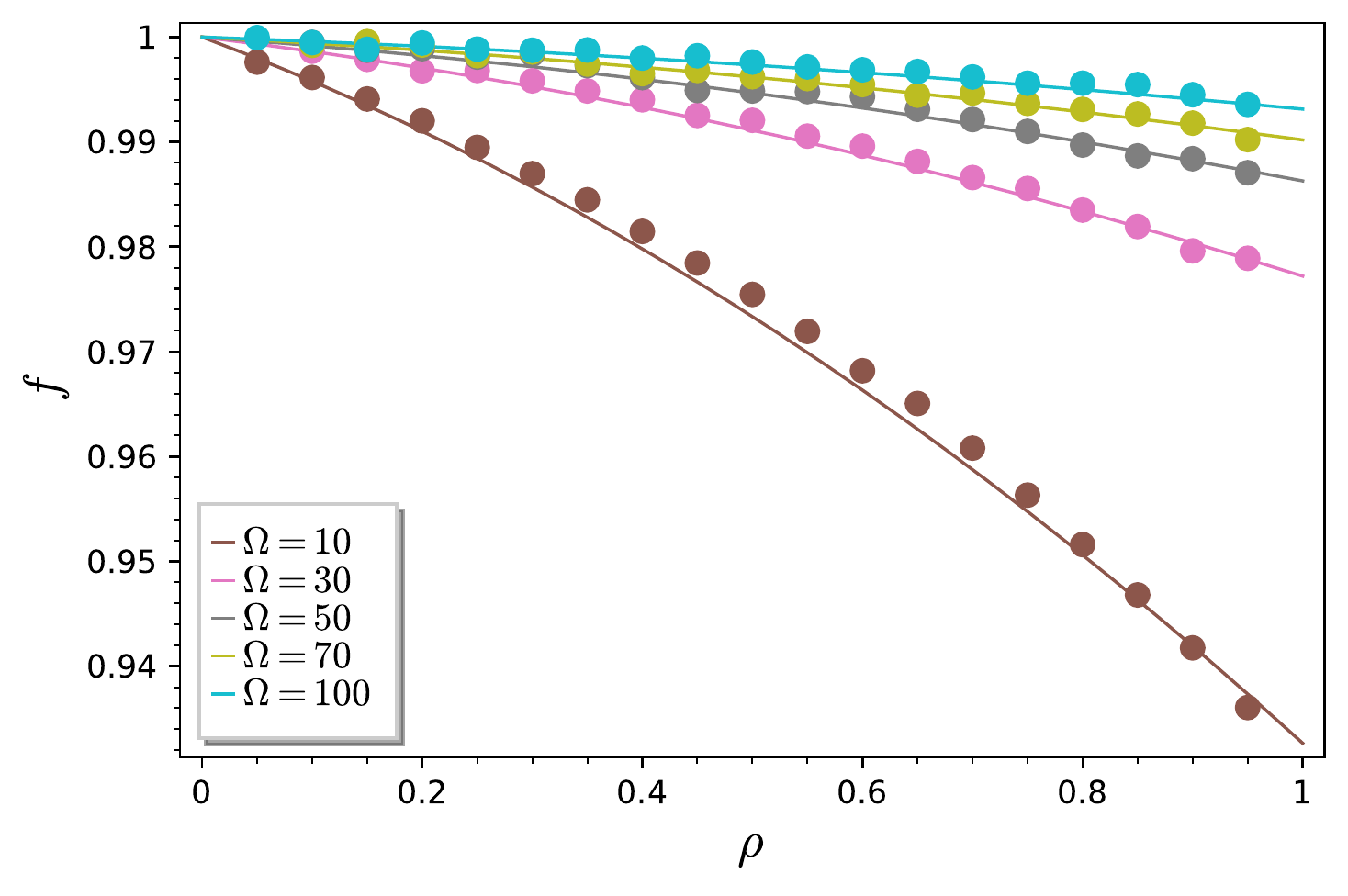}
	\caption{Correlation factor $f$ against density $\rho$ for soft-core interaction and different values of $\Omega$, the maximum number of particles allowed. Points are numerical results and curves correspond to Eq.\ \eqref{e.fsoft} with $c=0.156$. Black dotted and dashed curves correspond to analytical results of Nakazato \cite{nakazato} and Chaturvedi \cite{chaturvedi} respectively, obtained for $\Omega=1$.} \label{f.soft}
\end{figure}

Figure \ref{f.soft} shows the correlation factor against density for different values of $\Omega$; an adjusted value of $c=0.156$ was used. Numerical results (dots) were obtained from diffusion simulations in a two dimensional square lattice. Analytical expressions obtained by Nakazato \cite{nakazato} and Chaturvedi \cite{chaturvedi} for $\Omega=1$ are also shown for comparison, they are slightly larger than the numerical results. The same happens to our linear expression for $\Omega=1$, suggesting that higher order terms have to be included in the expansion of $R_{n_0,n_2}$ to obtain a more accurate approximation. Nevertheless, the expression \eqref{e.fsoft} satisfactorily matches numerical data for different values of $\Omega$ using the same value of $c$ in all cases. 

The expressions obtained by Nakazato and Chaturvedi for the two-dimensional square lattice and $\Omega=1$ are
\begin{align}
f_\text{Nakazato} &= \frac{(2-\rho)(1-\alpha)}{2-\rho-\alpha(2-3\rho)} \quad \text{with } \alpha=0.363 \nonumber\\
f_\text{Chaturvedi} &= 1 - \frac{2\rho}{(3-\beta)(2-\rho) + 2\rho} \quad \text{with } \beta=1.1894. \nonumber
\end{align}

\section{Extended hard core}
\label{s.extended}

\begin{figure}
	\includegraphics[width=\linewidth]{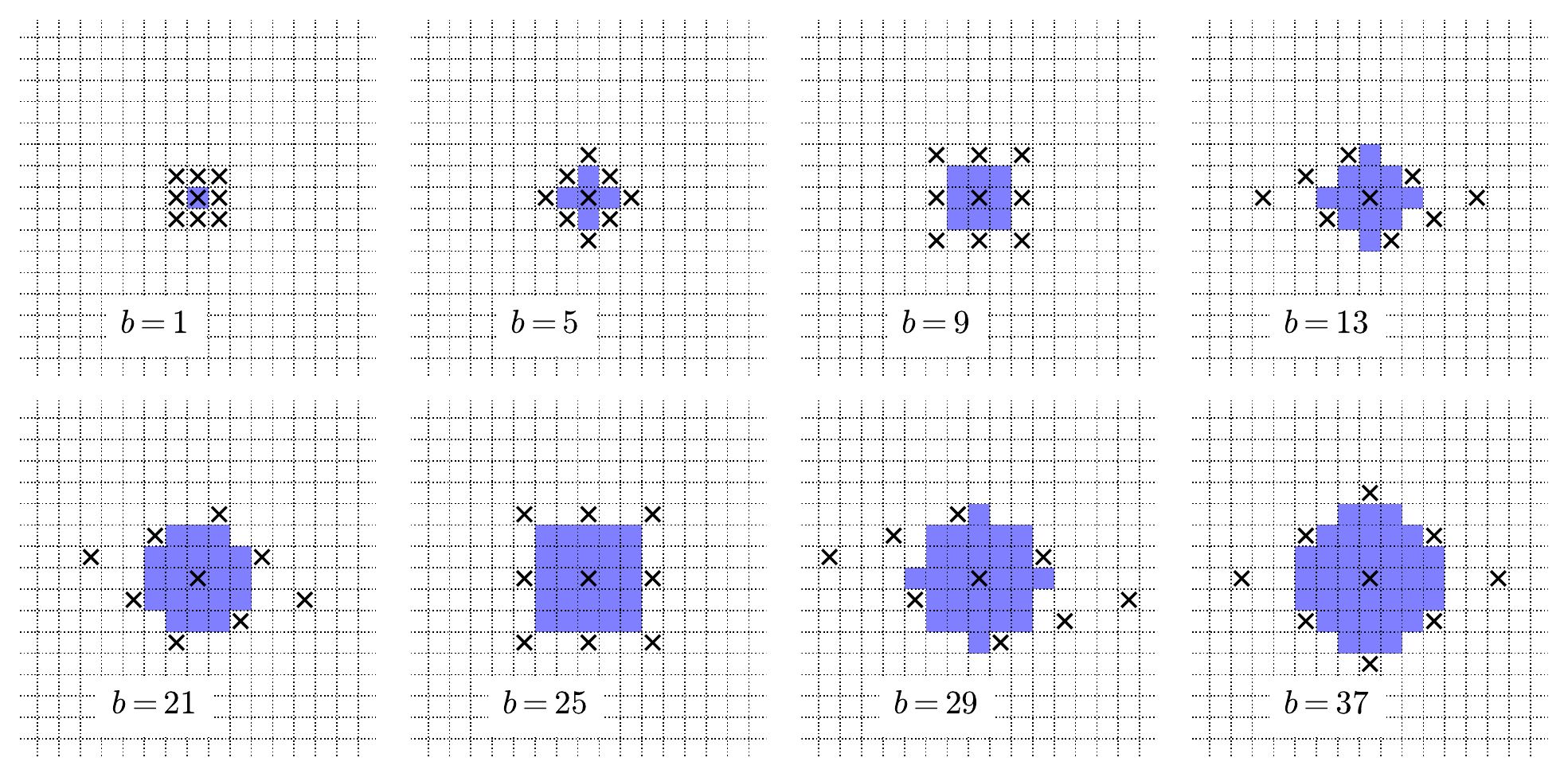}
	\caption{Exclusion regions, shown in blue around the central particle, for different values of $b$. A configuration with maximum concentration is partially represented in each case with crosses at the centers of the particles.} \label{f.ext}
\end{figure}

Diffusion of particles with extended hard core is analyzed in this section. Particles move in a two-dimensional square lattice; see \cite{marques} for a possible order-disorder phase transition in this kind of system. The center of a particle occupies a lattice site and there is an exclusion region around it that the center of other particles cannot occupy; some examples are shown in Fig.\ \ref{f.ext}.  A particle's center jumps randomly to neighboring sites, and the jump is allowed if the destination site does not belong to the exclusion region of another particle. Let us call $b$ the number of sites (or the area) of the exclusion region. If $b=1$ we have the situation of the previous section (with $\Omega=1$). The next case is $b=5$, where we have the center plus four nearest neighbors; by including the next-nearest neighbors we have $b=9$, a region of $3\times 3$ sites, etc. Hard disks are obtained in the limit of large $b$. Then, $b$ is not the particle size but the area of the exclusion region. For hard disks, the exclusion region has an area $\pi d^2$, with $d$ the particle diameter, while the particle area is $\pi d^2/4$. In analogy to hard disks, we define the packing fraction as $\xi = \rho b/4$ for $b>1$; and $\xi=\rho$ for $b=1$.

Knowing that the thermodynamic factor is $\Gamma = 1/(1-\rho)$ for $b=1$, we assume that for other values of $b$ it is approximately given by 
\begin{equation}\label{e.gammaext}
\Gamma = \frac{1}{1-\rho/\rho_\text{max}},
\end{equation}
where $\rho_\text{max}$ is the maximum possible value of concentration, given by the average number of particles per lattice site. Fig.\ \ref{f.ext} shows different shapes of the exclusion region around one particle for increasing values of $b$; the sequence starts from $b=1$ and a layer of nearest neighbors is added in each step. The crosses represent an example of a configuration with maximum concentration in each case, from which the value of $\rho_\text{max}$ is obtained; the values are given in the following table: 

\begin{center}
	\begin{tabular}{c|c|c|c|c}
	\rule[-1ex]{0pt}{2.5ex} $b$ & 1 & 5 & 9 & 13 \\
	\rule[-1ex]{0pt}{2.5ex} $\rho_\text{max}$ & 1 & 1/2 & 1/4 & 1/5 \\
	\hline
	\rule[-1ex]{0pt}{2.5ex} $b$ & 21 & 25 & 29 & 37 \\
	\rule[-1ex]{0pt}{2.5ex} $\rho_\text{max}$ & 1/8 & 1/9 & 1/10 & 1/12 \\
\end{tabular}
\end{center}

The maximum packing fraction is immediately obtained from $\xi_\text{max}=\rho_\text{max} b/4$ for $b>1$, and $\xi_\text{max}=1$ for $b=1$.

Considering transitions between cells of size $m\times m$, where $m$ is approximately equal to the particle diameter, the analysis of the previous section can be applied, resulting a correlation factor that has the concentration dependence given by Eq.\ \eqref{e.fsoft}, that is
\begin{equation}\label{e.fextended}
f = 1 - c_1 \xi + c_2\xi^2,
\end{equation}
where $c_1$ and $c_2$ are adjustable parameters. Using Eq. \eqref{e.Df2}, the tracer diffusivity is 
\begin{equation}\label{e.Dext}
D/D_0 = f/\Gamma = (1 - c_1 \xi + c_2\xi^2)(1-\xi/\xi_\text{max}).
\end{equation}

\begin{figure}
	\includegraphics[width=\linewidth]{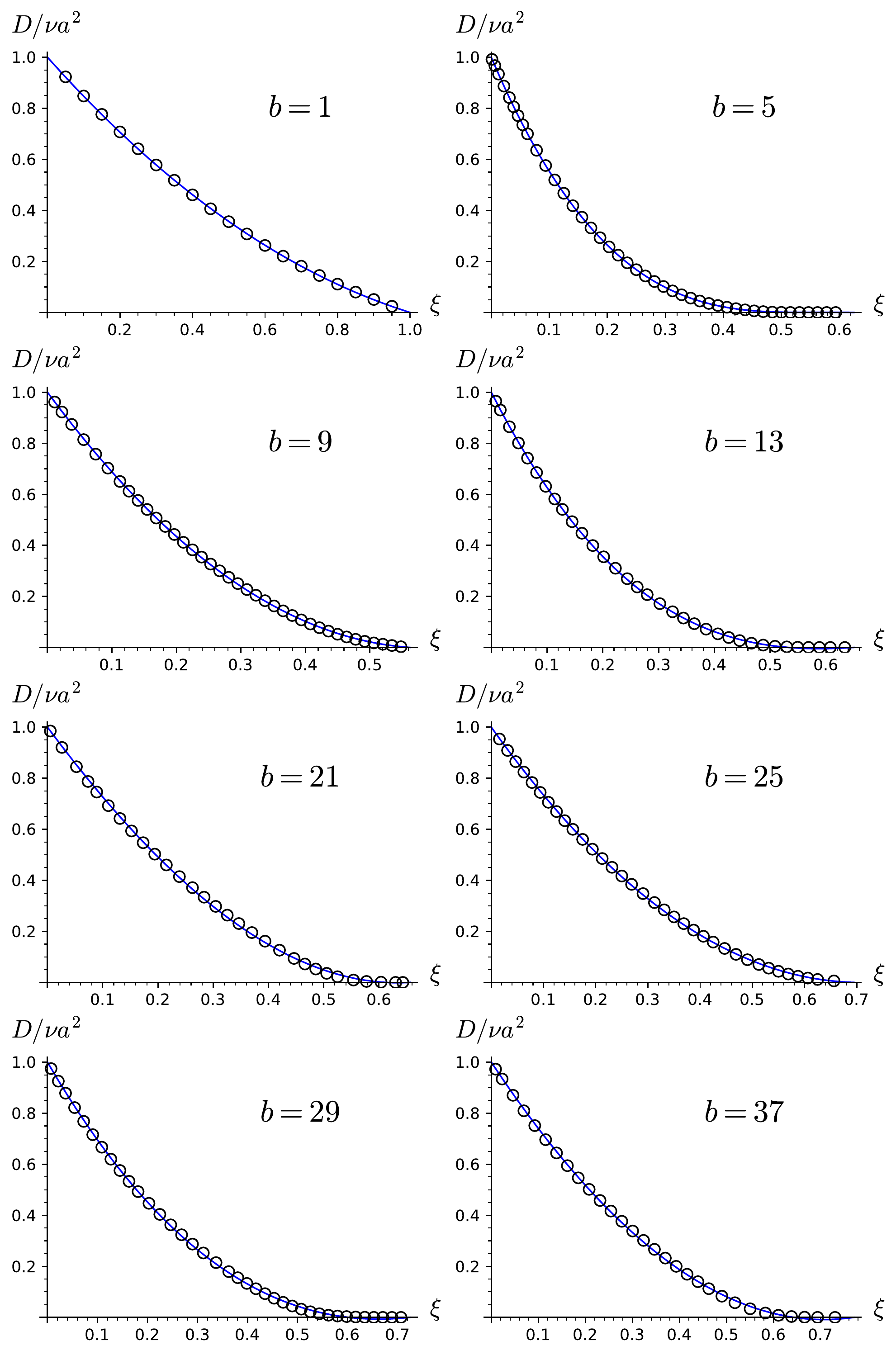}
	\caption{Normalized tracer diffusivity, $D/\nu a^2$, against packing fraction, $\xi$, for different values of $b$. Dots are numerical results and curves represent Eq.\ \eqref{e.Dext} with adjusted values of $c_1$ and $c_2$.}
	\label{f.Dext}
\end{figure}

Fig.\ \ref{f.Dext} shows numerical values of the tracer diffusivity, $D$, against packing fraction, $\xi$, for different values of exclusion region size, $b$. The results are well represented by Eq.\ \eqref{e.Dext} adjusting the values of parameters $c_1$ and $c_2$; the resulting parameters are shown in table \ref{tabla}.

\begin{table}
	\centering
\begin{tabular}{ccc}
	\rule[-1ex]{0pt}{2.5ex}  $b$ & $c_1$ & $c_2$ \\
	\hline
	\rule[-1ex]{0pt}{2.5ex} 1 & 0.593 & 0.0396 \\
	\rule[-1ex]{0pt}{2.5ex} 5 & 3.77 & 3.57 \\
	\rule[-1ex]{0pt}{2.5ex} 9 & 1.64 & 0.062 \\
	\rule[-1ex]{0pt}{2.5ex} 13 & 2.76 & 1.66 \\
	\rule[-1ex]{0pt}{2.5ex} 21 & 1.38 & -0.41 \\
	\rule[-1ex]{0pt}{2.5ex} 25 & 1.43 & 0.068 \\
	\rule[-1ex]{0pt}{2.5ex} 29 & 1.99 & 0.55 \\
	\rule[-1ex]{0pt}{2.5ex} 37 & 1.48 & -0.11 \\
\end{tabular}
\caption{Adjusted values of $c_1$ and $c_2$ used in Fig.\ \ref{f.Dext} for different values of $b$.}
\label{tabla}
\end{table}

\section{Summary and conclusions}
\label{s.conclusions}

In this manuscript we address the problem of the spatial correlation
effects when studying the tracer diffusivity on a regular lattice. We
have found a general expression for the correlation factor $f$, which
takes into account the memory effects of consecutive jumps. Interactions are represented macroscopically by the
excess chemical potential. Here we have analysed the cases of
hard-core, soft-core, and extended hard-core interactions. In all cases
the results from the Monte Carlo simulations show a good agreement
with the results predicted by the theoretical expression of $f$. As
expected, we note that for soft core the correlation effects decrease ($f$ tends to 1) when $\Omega$, the number of possible configurations for one particle within a lattice site, increases. This is because when $\Omega$ is large, a jump of a particle represents a minor change in the origin site. Thus, in equilibrium,
the particle will move with almost equal probability to any of its 4
neighboring sites in its next jump.

A minor but evident drawback of the theoretical approach is that it does not
yield a complete expression for $f$, since it contains one
free-parameter that needs to be adjusted for soft-core interactions, or two for extended hard-core. Nevertheless, it has the virtue of being a general expression that can be employed to study tracer diffusion in several systems ruled by different interactions. As we have shown
throughout these lines, memory effects play an important role in the
diffusivity of a tracer particle. A theoretical
understanding of the spatial correlations in a system of interacting
particles is key to gain a full comprehension of the nature of
transport processes.



\section*{Acknowledgments}
This work was partially supported by Consejo Nacional de Investigaciones
Cient\'ificas y T\'ecnicas (CONICET, Argentina, PUE 22920200100016CO).

\bibliographystyle{elsarticle-num}

\end{document}